\documentclass [11pt]{article}
\usepackage{amsmath}
\usepackage{amssymb}
\usepackage{amscd}
\usepackage{amsthm}
\usepackage{amsopn}
\usepackage{xspace}
\usepackage{verbatim}
\usepackage{amsmath}
\usepackage{amsfonts}
\newtheorem{theorem}{Theorem}
\newtheorem{lemma}{Lemma}[section]

\newtheorem{definition}{Definition}[section]
\newtheorem{acknowledgment*}{Acknowledgment}

\numberwithin{equation}{section}

\newcommand{\be}{\begin{equation}}
\newcommand{\ee}{\end{equation}}
\newcommand{\bd}{\begin{displaymath}}
\newcommand{\ed}{\end{displaymath}}

\newcommand{\R}{\mathbb R}

\renewcommand{\vec}[1]{\boldsymbol{#1}}

\begin{document}
\Large
\begin{center} {\bf On the mobility and efficiency of  mechanical systems }
\end{center}

\normalsize


\vskip 0.2cm

\begin{center}   Gershon Wolansky
 \\ Department of Mathematics, Technion, Haifa 32000, Israel%
\end{center}

\vskip 0.5cm


\vskip 0.5cm

\begin{abstract}
The definition of a mobilized system  and its efficiency are
introduced. The existence of an optimal (maximally efficient)
system is proved by an application of Young measures and
compensated compactness.

\end{abstract}
\section{Introduction}
\subsection{Motivation}
\vskip 0.5cm \noindent Which one is the exception- a motor boat,
 a  car,
 an airplane or a
 submarine?
 The immediate (perhaps, after a short
contemplation) answer is, of course, the car. Indeed, a car is the
only vehicle  in the above list which "defies" the law of
conservation of linear momentum. The others  create an opposite
stream in the medium in which they move (air, water) which, by
conservation of linear momentum, must be compensated by the motion
of the vehicle. Take away the friction of these vehicles and the
medium, and they will perform better (faster, more efficient).\par
The principle behind the motion of a car is different. It is
moving (that is, shifting from rest to cruise velocity) {\it
because} of the friction created by the contact of its tires with
the road. Take away the friction and the car will not be able to
move at all, independently of how hard you push the gas pedal. You
will not be able to stop either, if your friend, unwisely
attempting to help, gave you an initial push on a frictionless
road.
\par
So, what  is a car? If we strip it off the non-essential
components (radio, GPS, the fashioned  seat covers etc.), it is a
collection of components which can move with respect to each other
 under the preassign constraints caused by the mechanical structure.
  Unless you change gear or the
pressure you apply on the gas pedal, the motion of these inner
part can be assumed to be   periodic (again, with respect to the
frame of reference of the car itself). This concept has a lot in
common with the subject of optimal locomotion of a swimmer  in a
Stokes flow - where the motion is due to a periodic change of
shape  of the swimmer in the absence of inertia. The description
below is, in a sense, the mechanical analog of the swimmer model,
see [AGK]  and the references listed there. Another series of
publications which seems to be related to the present discussion
concerns molecular motors and the flashing rachet. See [CHK],
[CKK], DKK] and ref. therein.
\par
Note that in the case of micro-swimmers and molecular motors, it
seems that there is no intuitive way to predict the direction and
velocity of the swimmer (res. motor) from  its periodic motion. A
car, at a first glance, is different. However, the abstraction of
a car which we consider in this paper (and call a "mechanical
system") is, in a way, a generalization of the concept which
contains cars, microswimers, molecular motors and, perhaps, many
other objects whose dynamics are not inertial in nature.
\subsection{Objectives and outline of results}
A mechanical system is represented by a Lagrangian
$L=L(\dot{x},t)$. It is $T-$periodic in the time $t$, and $x(t)$
stands for the position of a reference point (say, the center of
mass of the system). Such a system is called "mobilized" if the
global  minimizer  $x(t)$ of the action $\int_0^T L(\dot{x},t)dt$
is {\it not} periodic, i.e $x(T)\not= x(0)$. In section~\ref{dec}
we attempt to justify  this model  and explain the reason why the
global minimum of the action represents the asymptotic motion of
the system under friction. In addition, we introduce  a reasonable
definition of efficiency, in terms of a relation involving the
speed $\overline{v}:= |x(T)-x(0)|/T$ due to the  action minimizer,
 and the minimal action $\overline{D}=\int_0^T L(\dot{x},t)dt$
itself. The number $\overline{D}$ stands for some indication of
the energy dissipated (or invested) per period, so a more
efficient system means larger $\overline{v}$ and smaller
$\overline{D}$. We scale the efficiency function $e_L$ of a
mechanical system $L$ in such a way that $0<e_L < 1$. Then, we ask
if either there is a (theoretical) possibility to achieve the
"most efficient" system $\overline{L}$ whose efficiency
$\overline{e}:= e_{\overline{L}}<1$, or  there is a sequence of
systems $L_n$ whose efficiencies $e_{L_n}\rightarrow
\overline{e}\leq 1$, but the "ideal", most efficient system
$\overline{L}$ does not exist.
\par
In section~\ref{homo} we consider a special case of mechanical
systems, where $L$ is a homogeneous function of $\dot{x}$, and
prove the first alternative: There {\it is} a "best" (most
efficient) mechanical system and its efficiency $\overline{e}$ is
always smaller than $1$.
\section{Description of the  model}\label{dec} Let us attempt to build a mathematical
 caricature the "car", composed of a finite number of "parts". To wit, assume
it is composed of a collection of $N$ points of respective masses
$m_i$ executing  orbits $x_i=x_i(t)$ on the line $ \mathbb{R}$,
$i=1,\ldots n$, so that
\begin{equation}\label{sumxi} x_i(t+T)=x_i(t)+ T \overline{v}\end{equation} where
$\overline{v}$ is the effective velocity of the car (with respect
to a reference frame attached to the road) and $T$ is the period
of one cycle.
\par
We assume  $\sum_1^N m_i =1$. The orbit $x_i(t)$ is given with
respect to a {\it fixed} frame of reference (attached to the
road). We may write it as
$$ x_i(t)=y_i(t) + x(t)$$ where $y_i$ is the orbit of the
corresponding point with respect to a reference attached to the
car and $x=x(t)$ is the orbit of the car as a whole with respect
to a reference attached to the road. In this representation $y_i$
is a periodic orbit, representing the  motion of a part forced by,
say,  the internal combustion of the engine.
\par
The simplest model for the motion of $x(t)$ is the linear forced
system \begin{equation}\label{lin} \ddot{x}+ \beta\dot{x} =
F(t)\end{equation} where $\beta>0$ is the friction coefficient and
$F$ is the total forcing acted on the car by the motion of its
inner parts. It is the sum of the forces $F:=\sum_i^N f_i$, where
$f_i$, the force applied  by the $i$ part,  is {\it defined} in
terms of $y_i$ as: \begin{equation}\label{linex} f_i(t)=
-m_i(\ddot{y}_i + \beta\dot{y}_i) \ . \end{equation} Now, define
\begin{equation}\label{calL}{\cal L}^{\{y\}}(\dot{x},t):=  \sum_{i=1}^N m_i
L(\dot{y}_i(t)+\dot{x}) \ .
\end{equation}
Then (\ref{lin}, \ref{linex}) can be summarized as the
Euler-Lagrange equation corresponding to the Lagrangian
\begin{equation}\label{calLbeta}{\cal L}_\beta^{\{y\}}(\dot{x},t):=e^{\beta t}{\cal
L}^{\{y\}}(\dot{x},t)\end{equation} where
\begin{equation}\label{sqL} L(s):= |s|^2/2 \ . \end{equation}
Indeed,  the Euler-Lagrange equation associated with ${\cal
L}_\beta^{\{y\}}$ under (\ref{sqL}) is
$$ 0=e^{\beta t}\left[ \sum_1^N m_i(\ddot{x}+\ddot{y}_i) + \beta
\sum_1^N m_i(\dot{x}+\dot{y}_i)\right] $$ which implies
(\ref{lin}) via (\ref{linex}) and the condition $\sum m_i=1$.
\par
The energy dissipated per cycle for an orbit $y_1, \ldots y_N, x$
is given by $\beta D$ where \begin{equation}\label{D} D:=
\sum_{i=1}^N m_i\int_0^T \frac{|\dot{y}_i+\dot{x}|^2 }{2}dt=
\int_0^T {\cal L}^{\{y\}}(\dot{x},t) dt \ . \end{equation} We now
generalize (\ref{calL}-\ref{D}) into:
\begin{definition} \label{L} A mechanical system  ${\bf L}_\beta$ is determined by a forced orbit
composed of $N$ periodic functions $\vec{y}= \{ y_i(t), \ldots
y_N(t)\}$  in terms of the Lagrangian ${\cal L}_\beta^{\{ y\}}$ as
given in (\ref{calLbeta}) and a convex function $L$ generalizing
(\ref{sqL}).
\end{definition}

 The moment associated with ${\cal L}^{\{y\}}$
is denoted by $ p={\cal L}^{\{y\}}_{\dot{x}}$. The Euler-Lagrange
equation associated with (\ref{calLbeta}) is
$$ \dot{p} + \beta p = 0 \Longrightarrow p(t)\rightarrow 0$$
so the asymptotic motion of a mechanical system ${\bf L}_\beta$ is
determined by the orbit $p(t)\equiv 0$.
\par
To elaborate, let
\begin{equation}\label{calH} {\cal H}^{\{y\}}(p,t)=\sup_{\zeta} \left[ p\cdot\zeta-{\cal
L}^{\{y\}}(\zeta,t)\right] \ .
\end{equation}
be the Hamiltonian associated with the Lagrangian ${\cal L}^{\{
y\}}$. The equation of motion corresponding to the non-dissipative
dynamics is given by
$$ \dot{x}(t)= {\cal H}^{\{y\}}_p(\lambda, t)$$
where $\lambda$ stands for the constant momentum $p$. If a
friction is applied, then $x=x(t)$ is determined by $p=0$, namely
$$ \dot{x}(t)={\cal H}^{\{y\}}_p(0, t)$$
is the asymptotic  motion of the system. It is a {\it global}
minimizer of the action determined by the Lagrangian ${\cal
L}^{\{y\}}$. Note  that
$$ \min_{x=x(t)} \int_0^T {\cal L}^{\{y\}}(\dot{x},t)dt = -\int_0^T {\cal
H}^{\{y\}}(0,t)dt \ . $$
 Let
$$\overline{v}(0)=\frac{1}{T}\int_0^T {\cal H}^{\{y\}}_p(0,t)dt\equiv \frac{x(T)-x(0)}{T}  \
, $$ where $x(t)$ is the global minimizer of the action.
\begin{definition} \label{v} A mechanical system for which $\overline{v}(0)\not=
0$ is called a {\it mobilized} system.
\end{definition}
\par\noindent The first result is somewhat disappointing:
\begin{theorem}\label{thm1} If $L$ is a quadratic function (\ref{sqL}), then
the system is not mobilized\ .\end{theorem} \begin{proof}
$$ {\cal L}^{\{y\}}(\dot{x},t)=\frac{1}{2}\sum_{i=1}^N
m_i\left|\dot{y}_i(t)+\dot{x}\right|^2\Longrightarrow {\cal
L}^{\{y\}}_{\dot{x}}=\sum_{i=1}^N
m_i\left(\dot{y}_i(t)+\dot{x}\right) \ . $$ In particular, ${\cal
L}^{\{y\}}_{\dot{x}}=0$ implies $\dot{x}=-\sum_{i=1}^N
m_i\dot{y}_i$. Since $y_i$ are periodic by definition, then
$T^{-1}\int_0^T\dot{x}dt=\overline{v}(0)=0$.
\end{proof}
We now generalize the energy dissipation in the linear case
(\ref{D}) for general ${\bf L}_\beta$ mechanical systems. We shall
denote by $\overline{D}$ the minimal action, and refer to it as
the {\it energy dissipated along a cycle}:
\begin{equation}\label{Ddef} \overline{D}=\min_{x=x(t)}\frac{1}{T}
\int_0^T{\cal L}^{\{y\}}(\dot{x},t)dt = -\frac{1}{T}\int_0^T {\cal
H}^{\{y\}}(0,t)dt \ . \end{equation}
\par
Next, we define the efficiency   of a mobilized system $\vec{y} :=
y_1, \ldots y_N$. This should indicate  the ratio of the distance
transversed per cycle to the dissipated energy. The right scaling
for this turns out to be \begin{equation}\label{eff} e_L(y_1,
\ldots y_N):=\frac{L\left(\overline{v}(0)\right)}{\overline{D}}=
-\frac{L\left( \frac{1}{T}\int_0^T {\cal
H}^{\{y\}}_p(0,t)dt\right)}{T^{-1}\int_0^T {\cal
H}^{\{y\}}(0,t)dt} \ .
\end{equation}
In fact, it can be proven that
\begin{lemma}\label{jensen} If $L$
is a convex function, then for any mechanical system composed of
$N$ periodic orbits  $y_i(t)= y_i(T+t), i=1,\ldots N$ , $$ 0<
e_L(y_1, \ldots y_N)\leq 1 \ . $$ If, moreover, $L$ is strictly
convex, then $e_L <1$.
\end{lemma}

\begin{proof} Let $x(t)$ be the {\it any} orbit. Then, $$
T^{-1}\int_0^T {\cal L}^{\{y\}}(\dot{x},t)dt = T^{-1}\sum_{i=1}^N
m_i\int_0^T L(\dot{y}_i(t) +\dot{x}(t))dt \ . $$ By Jensen's
inequality, the normalization condition $\sum m_i=1$, the
periodicity of $y_i$
 and the convexity of $L$: \begin{equation}\label{ineq}
T^{-1}\sum_{i=1}^N m_i \int_0^TL(\dot{y}_i(t) +\dot{x}(t))\geq
L\left(T^{-1}\int_0^T \sum_i^N
m_i(\dot{y}_i+\dot{x})dt\right)=L\left(T^{-1}\int_0^T
\dot{x}dt\right)  \ , \end{equation} so $0 < e_L \leq 1$. Now, if
$L$ is strictly convex then  the equality in (\ref{ineq}) holds if
and only if $N=1$ or $y_i\equiv y_j$ for all $1\leq i,j \leq N$.
But, in the later cases, the optimal orbit $x$ is evidently equal
to any component $y_i$, so it is a periodic function and
$\overline{v}(0)=0$. Hence the system is not mobilized and
efficiency is not defined (or $e_L=0$ altogether).
\end{proof}
\par
  Assume now
that a convex  Lagrangian $L$ is given, as well as $N\in
\mathbb{N}$ and $\{ m_1, \ldots m_N\}\in \mathbb{R}^{+,N}$,
$\sum_1^N m_i=1$. Let
\begin{equation}\label{lambdam}\Lambda_L := \left\{\vec{y}=( y_1, \ldots y_N)\in
C^1([0,1]; \mathbb{R}^N) \ ; \vec{y}(0)=\vec{y}(1)  \ .
\right\}\end{equation} Let
\begin{equation}\label{ebarold}\overline{e}_L:= \sup_{\vec{y}\in
\Lambda}e_L(\vec{y}) \ .
 \end{equation} We know that $\overline{e}_L\leq 1$.
 The
intriguing questions are

\begin{description}
\item{i)} \
Is $\overline{e}_L=1$?
\item{ii)} If $\overline{e}_L<1$, can the supremum in (\ref{ebarold}) be achieved in some
sense?
\end{description}
We try to answer these questions in the special case of
homogeneous Lagrangians.
\section{Homogeneous Lagrangians}\label{homo}
\par To fix the idea, let us concentrate on the case
$L(s)=|s|^\sigma$ for some $\sigma >1$. We shall further assume  a
unit period $T=1$. \par Let
\begin{equation}\label{Hsigma} H^\sigma:= \left\{ x=x(t): [0,1]\rightarrow \R \ , \
0\leq t\leq 1\ ; \ \int_0^1\left|\dot{x}\right|^\sigma <
\infty\right\} \ . \end{equation} Let also
\begin{equation}\Lambda_\sigma:= \left\{ \vec{y}\in \left(H^\sigma\right)^N \ , \vec{y}(0)=\vec{y}(1)
 \right\} \ . \end{equation}
 Given $m_i=m\in (0,1)$, $\sum_1^N m_i=1$ and $\vec{y}= (y_1, \ldots y_N)\in
 \Lambda_\sigma$
consider the Lagrangian
\begin{equation}\label{calL1}{\cal
L}(\dot{x},\dot{\vec{y}}):=\sum_{i=1}^N
m_i|\dot{y}_i+\dot{x}|^\sigma \ .
\end{equation}

Define also for any such $\vec{y}$ and $p\in\R^N$,
\begin{equation}\label{calH1} {\cal H}(p, \dot{\vec{y}}):=\sup_{\xi\in \R^N}\left\{ p\cdot\xi-{\cal
L}(\xi,\dot{\vec{y}}) \right\} \ ,
\end{equation} and
\begin{equation}\label{calH2} {\bf H}^{\vec{y}}(p):=\int_0^1{\cal
H}(p,\dot{\vec{y}}(t))dt
\end{equation}
Also, define: \begin{equation}\label{calH3} u(\dot{\vec{y}}):=
\left.\frac{\partial {\cal H}(p, \dot{\vec{y}})}{\partial
p}\right|_{p=0}\end{equation} and \begin{equation}\label{calH4}
<u>_{\vec{y}}:= \int_0^1u(\dot{\vec{y}}(t))dt \ . \end{equation}
\par We obtain via (\ref{calH1}-\ref{calH4}), (\ref{eff})  and
(\ref{ebarold}) \begin{equation}\label{esigma}
e_\sigma(\vec{y}):=-\frac{\left|<u>_{\vec{y}}\right|^\sigma}{{\bf
H}^{\vec{y}}(0)} \ \ \ ; \ \ \ \overline{e}_\sigma =
\sup_{\vec{y}\in \Lambda_\sigma}e_\sigma(\vec{y}) \ .
\end{equation}
One may wonder if, under the homogeneity condition, the mechanical
system is mobilized, i.e $\overline{e}_\sigma > 0$. The first
result we claim  is that this condition is equivalent to a
property of the function $u$ as defined in (\ref{calH3}), namely
\begin{lemma} The homogeneous system is mobilized if and only if
$u$ is not a linear function.
\end{lemma}
\begin{proof} If $u$ is a linear function, then for any
$\vec{y}\in \Lambda_\sigma$, $\int u\left(\dot{\vec{y}}\right)dt
=0$ by definition of $\Lambda_\sigma$. Indeed, the condition
$\vec{y}(0)=\vec{y}(1)$ is equivalent to
$\int_0^1\vec{\dot{y}}(t)dt=0$. \par Conversely, if $u$ is not a
linear function, then there exists $\vec{y}\in \Lambda_\sigma$ for
which $\int_0^1 u\left(\dot{\vec{y}}\right) dt \not= 0$. This
implies $e_\sigma(\vec{y}) > 0$ and, in particular, the system is
mobilized.
\end{proof}
The condition of non-linearity of  $u$ is rather delicate. In
fact,
\begin{lemma}\label{negative}
 If either $\sigma=2$ or $N=2$ then $u$ is a linear function.
\end{lemma}
\begin{proof}
 We already know that
$\sigma=2$ (quadratic Lagrangian) is not mobilized for any $N\in
\mathbb{N}$ by Theorem~\ref{thm1}. \par By the homogeneity of $L$
and the definition of $u$ we observe that $u$ satisfies the pair
of symmetries: $$ \forall \zeta\in \R^N \ , \ \vec{e}= (1, \ldots
1) \ ,\ and \ \lambda\in \R \ \ \ , \ \ \
 u(\vec{\zeta}+\lambda \vec{e})= u(\vec{\zeta})+\lambda $$
 \begin{equation}
 \forall \alpha\in \R^{+} \ , \
   \ u(\alpha \vec{\zeta})=\alpha u(\vec{\zeta}) \label{symmetry}
\end{equation}
 We conclude, therefore,
that $u(y_1,y_2)= f(y_1-y_2) + y_1$ holds for some function $f$ of
a single variable. From the second equality of (\ref{symmetry}) it
follows that
$$  f(\alpha (y_1-y_2)) + \alpha y_1 = \alpha f(y_1-y_2) +
\alpha y_1 \Longrightarrow \alpha f(\zeta) = f(\alpha\zeta) $$ for
any $\alpha\in \R$ and $\zeta\in \R$. Hence, $f$ is linear and so
is $u$.
\end{proof}
The main result of this paper is:
\subsection{Main result}
\begin{theorem}\label{main} If the homogeneous mechanical system (\ref{calL1}) is mobilized
 then there exists a maximizer  of $e_\sigma$ in the set
$\Lambda_\sigma$, and $\overline{e}_\sigma < 1$.
\end{theorem}
\par\noindent{\bf Remark:} \ Lemma~\ref{negative} implies that $N>2$  and  $\sigma\not=
2$ are necessary  for the condition of Theorem~\ref{main}. We
conjecture that these conditions are also sufficient. In any case,
it is not difficult to construct examples for mobilized
homogeneous systems.  For example, take $\sigma=3$, $N=3$ and
$m_1=m_2=m_3=1/3$. The function $u=u(y_1,y_2,y_3)$ can be readily
calculated as the root of a quadratic equation whose coefficients
are linear functions of $y_i$. The discriminant, however, is {\it
not} a complete square, so $u$ is not linear.
 \vskip .3in In the
rest of this section we prove Theorem~\ref{main}. \vskip
.2in\noindent
 From (\ref{symmetry}) we obtain that ${\cal H}(0,\dot{\vec{y}})$  is $\sigma-$homogeneous, that is
 $$ \alpha\in \mathbb{R} \ , \ \ {\cal H}(0,\alpha \dot{\vec{y}})=|\alpha|^\sigma
 {\cal H}(0,\dot{\vec{y}})  $$ as well as
 $$ <u>_{\alpha \vec{y}} = \alpha<u>_{\vec{y}} \ . $$
In particular
\begin{equation}\label{scale}
 e_\sigma(\alpha \vec{y})=e_\sigma(\vec{y}) \ , \ \forall\alpha
\in \mathbb{R}, \vec{y}\in \Lambda_\sigma  \ . \end{equation} In
addition, $e_{\sigma}(\vec{y})$ is clearly invariant under
diagonal shifts $\vec{y}\rightarrow \vec{y}(t)+ \beta(t)\vec{e}$
where $\vec{e}= ( 1, \ldots 1)\in \mathbb{R}^N$. Define now
 $$\Lambda_\sigma^0:= \left\{
\vec{y}= (y_1, \ldots y_N)\in \Lambda_\sigma \ ; \ \ y_1\equiv 0 \
. \right\} $$ and
\begin{equation}\label{sphere} S_\sigma=
\left\{ \vec{y}\in\Lambda^0_\sigma \ ; \ \
\int_0^1\left|\dot{\vec{y}}(t))\right|^\sigma dt = 1\right\} \ \ \
; \ \ \ B_\sigma= \left\{ \vec{y}\in\Lambda^0_\sigma \ ; \ \
\int_0^1\left|(\dot{\vec{y}}(t))\right|^\sigma dt \leq 1\right\} .
\end{equation}
By the scaling  (\ref{scale}) and the diagonal shift invariance we
conclude that
\begin{equation}\label{effdef} \overline{e}_\sigma:=\sup_{\vec{y}\in \Lambda_\sigma} e_\sigma(\vec{y}) =
\sup_{\vec{y}\in S_\sigma} e_\sigma(\vec{y})= \sup_{\vec{y}\in
B_\sigma- \{0\}} e_\sigma(\vec{y}) \ .
\end{equation}
Let now $\vec{y}_j$ be a maximizing sequence of $e_\sigma$ in
$S_\sigma$.  There is a weak limit $\vec{y}_\infty\in B_\sigma$ of
this sequence. The inequality
$$ \lim_{j\rightarrow\infty} {\bf H}^{\vec{y}_j}(0)\leq {\bf
H}^{\vec{y}_\infty}(0) \ $$ holds since ${\bf H}^{\vec{y}}(0)$ is
upper-semi-continuous, but we do not  have the same claim for
$<u>_{\vec{y}}$.  So, we cannot prove that $vec{y}_\infty$ is a
maximizer of $e_\sigma$.
\par
Another problem is that we may have $\vec{y}_\infty= \vec{0}$, so
$e_\sigma(\vec{y}_\infty)$ is not defined at all. As an example,
let $\vec{y}\in \Lambda_\sigma$ and assume that
$<u>_{\vec{y}}\not= 0$. This, in particular, implies that $x(t):=
\int^t u\left(\dot{\vec{y}}\right)= \tilde{x}(t) + \lambda t$
where $\tilde{x}$ is a periodic function and $\lambda\not= 0$. If
we replace $\vec{y}$ by $\vec{y}_j=\vec{y}_j(t):=
j^{-1}\vec{y}(jt)$ for $j\in \mathbb{N}$, using the periodicity of
$\vec{y}$ to define $\vec{y}_j$ on $(0,1)$,  the following claims
are straightforward:
\begin{description}
\item{a)} $\vec{y}_j\in\Lambda_\sigma$ for any $j\in \mathbb{N}$.
\item{b)}   $\lim_{j\rightarrow\infty} \vec{y}_j=0$ weakly.
\item{c)} $x_j=j^{-1}x(jt)= j^{-1}\tilde{x}(jt)+ \lambda t=
\int^t u\left(\dot{\vec{y}_j}\right)$.
 \item{d)}  $<u>_{\vec{y}_j}= \lambda$ for any $j\in \mathbb{N}$.
\end{description}
In particular, we find out  that $e_\sigma(\vec{y}_j)=
e_\sigma(\vec{y})$,  while $e_\sigma$ is not defined for the weak
limit $\lim_{j\rightarrow\infty} \vec{y}_j=\vec{0}$.
\subsection{Relaxation}
 To overcome the last difficulty we shall extend the
definition (\ref{calH2}-\ref{calH4}) as follows: Let $$ {\cal
P}^N:= \{ Probability \ Borel \ measures \ on \ \R^N \}
$$ Given  $\nu\in {\cal P}^N$, let
\begin{equation}\label{calH2nu} {\bf H}^{\nu}(p):=\int_{\R^N}{\cal
H}(p,v)\nu(dv)
\end{equation}
and \begin{equation}\label{calH4nu} <u>_{\nu}:= \int_{\R^N}
u(v)\nu(dv) \ .
\end{equation}
\par The generalization of (\ref{esigma}) is given by
\begin{equation}\label{esigmanu}
e_\sigma(\nu):=-\frac{\left|<u>_{\nu}\right|^\sigma}{{\bf
H}^{\nu}(0)}  = \frac{\left|<u>_\nu\right|^\sigma}{\int_{\R^n}
\sum_1^N m_i |v_i+u(\vec{v})|^\sigma\nu(d\vec{v})}\end{equation}
We shall further extend the definitions of $\Lambda_\sigma$ and
$\Lambda_\sigma^0$ as follows:
 \begin{equation}\label{overlambda}\overline{\Lambda}_\sigma:=
\left\{\nu=\nu(dv) \in {\cal P}^N \ \ ; \
\int_{\R^N}|v|^\sigma\nu(dv)< \infty \ \ \ ; \ \ \
\int_{\R^N}v_i\nu(dv)=0 \ ,  1\leq i\leq N \right\}\end{equation}
\begin{equation}\label{overlambda0}\overline{\Lambda}^0_\sigma:=
\left\{ \nu\in \overline{\Lambda}_\sigma \ ; \
\nu=\delta_{v_1}\mu(dv_2, \ldots dv_N) \ \ where\ \mu\in {\cal
P}^{N-1} \right\}\end{equation}
\begin{lemma}\label{monge}
For any $\vec{y}\in \Lambda_\sigma$ (res. $\vec{y}\in
\Lambda^0_\sigma$) there exists $\nu\in \overline{\Lambda}_\sigma$
(res. $\nu\in \overline{\Lambda}^0_\sigma$)   so that
$e_\sigma(\nu)=e_\sigma(\vec{y})$. Conversely, for any $\nu\in
\overline{\Lambda}_\sigma$ (res. $\nu\in
\overline{\Lambda}^0_\sigma$) there exists $\vec{y}\in
\Lambda_\sigma$ (res. $\vec{y}\in \Lambda^0_\sigma$) so that
 $e_\sigma(\nu)=e_\sigma(\vec{y})$.
\end{lemma}
\par\noindent {\bf Remark}: The measure $\nu$ associated with
$\vec{y}$ is related to {\it Young measure}. In general, however,
Young measures are used to study the oscillatory behavior of a
weak limit of $ \mathbb{L}^\infty$ sequences (see, e.g., [E]).
 \begin{proof} For the first part, define $\nu(dv)= \int_0^1 \prod_1^N
 \delta_{v_i-\dot{y}_i(t)}$. To elaborate, the measure $\nu$
 corresponding to $\vec{y}$ is obtained by its application on test
 functions $\phi\in C_0\left(\R^N\right)$:
 $$ \int_{\R^N}\phi(v)\nu(dv)= \int_0^1
 \phi\left(\dot{\vec{y}}(t)\right) dt \ . $$
 If $\vec{y}\in \Lambda_\sigma$ then the above equality also
 extend to $\phi(\vec{\dot{y}})=\vec{\dot{y}}$ and $\phi(\vec{\dot{y}})=
 |\vec{\dot{y}}|^\sigma$. In particular
  $$ \int_{\R^N}v\nu(dv)= \int_0^1
 \dot{\vec{y}}(t) dt=0 \ ; \ \ \int_{\R^N}|v|^\sigma\nu(dv)= \int_0^1
 \left|\dot{\vec{y}}\right|^\sigma(t) dt< \infty \ .$$
 Finally, the equalities \begin{equation}\label{identification}{\bf H}^{\vec{y}}(0)={\bf H}^{\nu}(0) \
 \ ; \ \ <u>_\nu=<u>_{\vec{y}}\end{equation}
 hold under this identification.
 \par
 For the second part we use
 Theorem 2.1 of [Am] to observe the following:
 \par\noindent
For any such $\nu$ there exists a Borel function
$T:[0,1]\rightarrow \R^N$ which push forward the Lebesgue measure
$dt$ on $[0,1]$ to $\nu$. That is, for any test function  $\phi\in
C_0\left(\R^N\right)$,
$$ \int_0^1\phi(T(t))dt= \int_{\R^N}
\phi(v)\nu(dv) \ . $$ In fact, Theorem 2.1 of [Am] claims the
equality between the infimum of Monge and the minimum of the
Kantorowich transport plan of probability measures $\mu_1$ to
$\mu_2$ on $ \mathbb{R}^N$, provided $\mu_1$  has no atoms. This
implies, in particular, that the set of Borel mappings
transporting $\mu_1$ to $\mu_2$ is not empty. In our case we use
this result under the identification of $\mu_1$ with the Lebesgue
measure on $[0,1]$ (considered as a Hausdorff measure in $\R^N$
for the embedded interval) and $\mu_2$ with $\nu$.
\par
So, set $\vec{y}(t):= \int^t T(s)ds$. Is is absolutely continues
function satisfying $\dot{\vec{y}}= T$ a.e. \ The definition
(\ref{overlambda}) implies also that $\int_0^1
\left|\dot{\vec{y}}\right|^\sigma = \int
|v|^\sigma\nu(dv)<\infty$, as well as $\int_0^1 \dot{\vec{y}} dt =
\int v\nu(dv) = 0$, which yields the periodicity of $\vec{y}$. The
equality (\ref{identification}) holds under this identification as
well.
 \end{proof}

\subsection{Proof of Theorem~\ref{main}}
Define now
\begin{equation}\label{psi} \psi(\vec{v}):= |u(\vec{v})| + \sum_{i=2}^N |v_i+u(\vec{v})|
\ ,  \ \vec{v}= (0, v_2, \ldots v_N) \ . \end{equation} We shall
summertime some properties of $\psi$ which will be needed later:
\begin{lemma}\label{psiprop} There exists $A>0$ so that \begin{equation}\label{psiest}
A^{-1}|\vec{v}|\leq |\psi(v)|\leq A |\vec{v}| \ \ \forall \vec{v}=
(0, v_2, \ldots v_N)  \ . \end{equation} In addition, for any
$\nu\in \overline{\Lambda}_\sigma^0$,
\begin{equation}\label{psiest1} A^{-1} \int_{\R^{N}} |\psi(v)|^\sigma\nu(dv) < -{\bf H}^\nu(0)  \leq \ A
\int_{\R^{N}} |\psi(v)|^\sigma \nu(dv).
\end{equation}
\end{lemma}
\begin{proof}
The estimate (\ref{psiest}) follows from the homogeneity of $u$,
namely $u(\alpha \vec{v})=\alpha u(\vec{v})$, as well as from the
evident property $u(0,v, \ldots  v)=0  \Longleftrightarrow v=0$.
The estimate (\ref{psiest1}) follows from (\ref{calL1},
\ref{calH1}) and (\ref{calH2nu}), as well as (\ref{psiest}).
\end{proof}
 Given $1<q<\sigma$, let
 $\overline{S}^q_\sigma$ be the unit sphere (res. $\overline{B}^q_\sigma$ the unit
ball) defined by \begin{equation}\label{oversphere}
\overline{S}^q_\sigma= \left\{ \nu\in\overline{\Lambda}^0_\sigma \
; \ \ \int_{\R^{N-1}} |\vec{v}|^q\nu(dv) = 1\right\} \ \ \ ; \ \ \
\overline{B}^q_\sigma=\left\{ \nu\in\overline{\Lambda}^0_\sigma \
; \ \ \int_{\R^{N-1}} |\vec{v}|^q\nu(dv) \leq 1\right\}
\end{equation}
The analogous of (\ref{effdef}) also holds due to the scaling
invariance (\ref{scale}):
\begin{lemma}\label{ebar} \ \ \ \ \ \ \
$ \overline{e}_\sigma= \sup_{\nu\in \overline{S}^q_\sigma}
e_\sigma(\nu) \ . $
\end{lemma}
\begin{proof}
We only have to show that any $\nu\in\overline{\Lambda}_\sigma$
(res. $\nu\in\overline{\Lambda}^0_\sigma$)  can be transformed
into $\hat{\nu}\in \overline{S}_\sigma$ (res. $\hat{\nu}\in
\overline{S}^0_\sigma$) so that
$e_\sigma(\nu)=e_\sigma(\hat{\nu})$. Let $\hat{\nu}(dv)= \beta^N
\nu(\beta dv)$ for $\beta >0$. The homogeneity properties
(\ref{scale}) imply that, indeed, $e_\sigma
(\hat{\nu})=e_\sigma(\nu)$ for any such $\beta$. In addition,
$\int v\hat{\nu}(dv) = \beta^{-1}\int v \nu(dv)=0$ if $\nu\in
\overline{\Lambda}_\sigma$. However, $\int|v|^q \hat{\nu}(dv)=
\beta^{-q}\int|v|^q \nu(dv)$, so $\hat{\nu}\in
\overline{S}_\sigma$ if $\beta= \left(\int
|v|^q\nu(dv)\right)^{1/q}$.
\end{proof}
Let $\nu_j$ be a maximizing sequence of $e_\sigma$ in
$\overline{S}_\sigma^0$.  Since the $q>1$ moments of $\nu_j$ are
uniformly bounded, this sequence is compact (tight)  in the weak
topology of measures. Let $\nu_\infty$ be the weak limit of
$\nu_j$. Since $|u(v)|\leq A|v|$ for some $A>0$ it follows that
the sequence of (signed) measures $u(v)\nu_j(dv)$ is tight as
well, and that
\begin{equation}\label{tightu}\lim_{j\rightarrow\infty} u(v)\nu_j(dv)= u(v)\nu_\infty(dv) \
. \end{equation} On the other hand, it is not a-priori evident
that $\nu_\infty\in \overline{S}_\sigma^0$, since the sequence
$|v|^q \nu_j$ is not necessarily tight, so we only know
\begin{equation}\label{vsupq} \int |v|^q\nu_\infty(dv) \leq 1 \ .
\end{equation} We claim, however, that, in fact, the $\sigma$
moments of $\nu_j$ are uniformly bounded. For, if
$\int|v|^\sigma\nu_j(dv)\rightarrow\infty$, then by (\ref{psiest},
\ref{psiest1}), also $-{\bf H}^{\nu_j}(0)\rightarrow\infty$. Since
we also now that $<u>_{\nu_j}$ are uniformly bounded, then
$$ e_\sigma(\nu_j) = -\frac{ <u>_{\nu_j}}{{\bf H}^{\nu_j}(0)}
\rightarrow 0 \ , $$ contradicting the assumption that $\nu_j$ is
a  maximum sequence for $e_\sigma$.
\par
Since $q<\sigma$ by assumption it follows that $|v|^q\nu_j$ is a
tight sequence as well, so there is, in fact, an equality in
(\ref{vsupq}). In particular, it follows that $\nu_\infty\not=
\delta_0$. Moreover, using (\ref{psiest}, \ref{psiest1}) for
$\nu_\infty$ and the equality in (\ref{vsupq}) again, we also have
$$ -{\bf H}^{\nu_\infty}(0) >0 \ . $$
On the other hand, $-{\bf H}^\nu(0)$ is nothing but the
expectation of $\nu$ with respect to a positive, continuous
function ${\cal L}(v)=\sum_{i=2}^N m_i |v_i-u(v)|^\sigma$. Hence
$$ -\lim_{j\rightarrow\infty}{\bf H}^{\nu_j}(0) := \lim_{j\rightarrow\infty}\int {\cal L}(v)\nu_j(dv)
\geq \int {\cal L}(v)\nu_\infty(dv) := -{\bf H}^{\nu_\infty}(0) \
. $$ This, together with (\ref{tightu}), implies that
\begin{equation}\label{final} e_\sigma (\nu_\infty) \geq \lim_{j\rightarrow\infty}
e_\sigma (\nu_j) \ . \end{equation} Again, $\nu_j$ is a maximizing
sequence for $e_\sigma$ hence there is an equality in
(\ref{final}), so $\overline{e}_\sigma= e_\sigma(\nu_\infty)$. By
the second part of  Lemma~\ref{monge} we obtain the existence of
$\vec{y}_\infty\in \Lambda_\sigma^0$ for which
$e_\sigma(\nu_\infty)= e_\sigma (\vec{y}_\infty)$.
 \ \ \  $\qed$
\begin{center} {\bf References}\end{center}
\begin{description}
\item{[Am]} Ambrosio, L, {\it Lecture notes on Pptimal Transport
problems}, Lect. Notes in Math., "Mathematical aspects of evolving
interfaces" (CIME seris Madeira (PT) 2000, {\bf 1812}, P. Colli
and J.F. Rodrigues, Eds., 1-52, 2003
\item{[AGK]} J.E. Avron, O ; Gat and O. Kenneth, {\it Optimal
swimming at low Reynolds numbers}, Phys. Rev. Lett., {\bf 93}, \#
18, 2004.
\item{[CHK]} Chipot, M ; Hastings, S and Kinderlehrer, D: {\it
Transport in a molecular system}, M2AN Math. Model. Num. Anal.,
{\bf 38} , \#6, 1011-1034, 2004
\item{[CKK]}  Chipot, M;   Kinderlehrer, D and Kowalczyk, M:
{\it A  variational principle for molecular motors, dedicated to
Piero Villagio on the occasion of his 60 birthday}, Mechanica {\bf
38}, \#5, 505-518, 2003
\item{[DKK]}  Dolbeault, J;   Kinderlehrer, D and Kowalczyk,
M:{\it Remarks about the flashing rachet, Partial differential
equations and inverse problems}, Contemp. Math. {\bf 362},
167-175,  2004
\item{[E]} Evans, L.C.: {\it Weak convergence methods for no
nonlinear partial differential equations}, CBMS, {\bf 74}, 1990
\item{[W]} Wolansky, G: {\it Rotation numbers for measure-valued
circle maps}, J. D'analyse, to appear.
\end{description}
\end{document}